\documentclass[]{elsarticle}

\usepackage{graphicx}
\usepackage{epsfig}		
\usepackage{dcolumn}
\usepackage{bm}
\usepackage{amsmath}
\usepackage{braket}
\usepackage{natbib}
\usepackage{color}
\usepackage{feynmf}
\usepackage{pifont}
\usepackage{geometry}
\usepackage{fleqn}
\usepackage{txfonts}
\usepackage{amssymb }

\biboptions{numbers,sort&compress}

\def\0{\mbox{\tiny $0$}}
\def\1{\mbox{\tiny $1$}}
\def\2{\mbox{\tiny $2$}}
\def\3{\mbox{\tiny $3$}}
\def\4{\mbox{\tiny $4$}}
\def\5{\mbox{\tiny $5$}}
\def\6{\mbox{\tiny $6$}}
\def\7{\mbox{\tiny $7$}}
\def\8{\mbox{\tiny $8$}}
\def\9{\mbox{\tiny $9$}}

\def\f14{\mbox{\tiny $\frac{1}{4}$}}

\journal{Physics Letters B}

\begin{document}

\title{Scalar field dark matter and the Higgs field}

\author[ob]{O. Bertolami}
\ead{orfeu.bertolami@fc.up.pt}
\author[ob]{Catarina Cosme}
\ead{catarinacosme@fc.up.pt}
\author[jr]{Jo\~{a}o G. Rosa}
\ead{joao.rosa@ua.pt}

\address[ob]{Departamento de F\'isica e Astronomia, Faculdade de Ci\^{e}ncias da
Universidade do Porto and  Centro de F\'isica do Porto, Rua do Campo Alegre 687, 4169-007, Porto, Portugal.}
\address[jr]{Departamento de F\'isica da Universidade de Aveiro and Center for Research and Development in Mathematics and Applications (CIDMA) Campus de Santiago, 3810-183 Aveiro, Portugal; also at Departamento de F\'isica e Astronomia, Faculdade de Ci\^{e}ncias da
Universidade do Porto, Rua do Campo Alegre 687, 4169-007, Porto, Portugal.}

\begin{abstract}
We discuss the possibility that dark matter corresponds to an oscillating scalar field coupled to the Higgs boson. We argue that the initial field amplitude should generically be of the order of the Hubble parameter during inflation, as a result of its quasi-de Sitter fluctuations. This implies that such a field may account for the present dark matter abundance for masses in the range $10^{-6} - 10^{-4}$ eV, if the tensor-to-scalar ratio is within the range of planned CMB experiments. We show that such mass values can naturally be obtained through either Planck-suppressed non-renormalizable interactions with the Higgs boson or, alternatively, through renormalizable interactions within the Randall-Sundrum scenario, where the dark matter scalar resides in the bulk of the warped extra-dimension and the Higgs is confined to the infrared brane.
\end{abstract}

\date{\today}
\maketitle



\section{Introduction} \label{intro}

The existence of a significantly undetected non-relativistic matter component in the Universe is widely accepted, with plenty of evidence arising from different sources. In particular, the flatness of the rotational curves of galaxies requires a significant dark matter component to account for the inferred dynamical galactic mass. In addition, the invisible mass of galaxy clusters and the temperature and polarization anisotropies of the Cosmic Microwave Background (CMB) radiation indicate a dark matter component that accounts for about 26\% of the present energy balance in the Universe.  The origin and the constitution of dark matter remain, however, unknown, despite the large number of candidates that arise in theories beyond the Standard Model of Particle Physics (see e.g.~\cite{Bertone:2004pz} for a review). 

The recent discovery of the Higgs boson at the Large Hadron Collider \cite{Aad:2012tfa, Chatrchyan:2012xdj} has opened up new possibilities for understanding the nature of dark matter. In fact, several works in the literature have already considered the possibility that dark matter interacts with the Higgs field in a variety of forms. The first models \cite{Silveira:1985rk, Bento:2000ah, Burgess:2000yq, Bento:2001yk, Seto:2001ju} considered an extension of the Standard Model with an additional singlet scalar field, $\phi$, with renormalizable interactions with the Higgs field of the form:
\begin{equation}
V\left(\phi,h\right)=\frac{m_{\phi}^{2}}{2}\phi^{2}+\frac{\lambda_{\phi}}{4}\phi^{4}+{1\over 2}g^2\phi^{2}h^{\dagger}h~,
\end{equation}
where $m_{\phi}$ is the field mass, $\lambda_{\phi}$ its self-coupling term and $g$ the coupling term between the ``phion"  and the Higgs field. Several analyses of this and related models have been performed in the literature \cite{Patt:2006fw, MarchRussell:2008yu, Biswas:2011td, Pospelov:2011yp, Mahajan:2012nc, Cline:2013gha, Kouvaris:2014uoa, Costa:2014qga, Duerr:2015aka, Enqvist:2014zqa, Tenkanen:2015nxa, Han:2015hda, Han:2015dua, Kainulainen:2016vzv} and this possibility has become widely known as ``Higgs-portal" dark matter. A connection to dark energy has also been suggested in \cite{Bertolami:2007wb, Bertolami:2012xn}. 

Most of the works in the literature focus, however, on dark matter candidates whose abundance is set by the standard decoupling and freeze-out mechanism, with masses in the GeV-TeV range. In this work, we consider an alternative possibility in which the scalar field $\phi$ acquires a large expectation value during inflation and begins oscillating after the electroweak phase transition, behaving as non-relativistic matter. Although a related scenario was considered e.g.~in Ref. \cite{Tenkanen:2015nxa}, in the latter case interactions are sufficiently large to lead to the decay of the scalar condensate and thermalization of the $\phi$-particles, so that the present-day dark matter abundance also corresponds to a GeV-TeV WIMP thermal relic. 

Our proposal considers a scenario where the dark matter field is part of a hidden/sequestered sector with an inherent conformal symmetry/scale invariance, which is broken only by feeble interactions with the Higgs boson. This implies that the field's mass and self-couplings are extremely small, which in particular leads to a long-lived oscillating scalar condensate that is never in thermal equilibrium in the cosmic history. We will consider particular models where this generic idea can be realized, and show that such a field can naturally account for the present dark matter abundance.

The article is organized as follows. In Sec.~\ref{sec:Oscillating scalar field as dark matter candidate} we describe the main properties of our scenario, focusing on the dynamics of an oscillating scalar field in the post-inflationary Universe and considering that the field only acquires mass through the Higgs mechanism after the electroweak phase transition. We determine, in particular, the relation between the field's mass and initial amplitude required in order to explain the observed dark matter abundance.  In Sec.~\ref{sec: constraints Infl} we discuss the dynamics of the field during inflation and compute the average field amplitude that results from quasi-de Sitter fluctuations and sets the initial conditions for the post-inflationary evolution. Finally, in Sec.~\ref{sec:Dark matter coupled to Higgs via non-renormalizable operators} and Sec. \ref{sec:Scalar field dark matter in a higher dimensional theory} we describe particular realizations of a weak coupling between the $\phi$ and Higgs fields leading to the required field mass to account for dark matter, considering firstly the case of non-renormalizable operators and secondly a bulk scalar field in the Randall-Sundrum scenario for a warped extra-dimension. We summarize our conclusions and prospects for future work in Sec.~\ref{concl}.
 


\section{Oscillating scalar field as dark matter} \label{sec:Oscillating scalar field as dark matter candidate}

Let us start by reviewing why a homogeneous oscillating field, $\phi$, with a potential dominated by a quadratic term, $V\left(\phi\right)=\frac{1}{2}m_{\phi}^{2}\phi^{2}$, behaves as non-relativistic matter. In a generic cosmological epoch where the scale factor evolves as $a\left(t\right)=(t/t_i)^p$, with $p>0$ and $a(t_i)=1$, the Hubble parameter is simply $H=p/t$. The field  $\phi$ then satisfies the Klein-Gordon equation:
\begin{equation}
\ddot{\phi}+3{p\over t}\dot{\phi}+m_{\phi}^{2}\phi=0~.\label{eq:KG generic}
\end{equation}
For $m_{\phi}t\gg1$, the solution of this equation is then approximately given by:
\begin{equation}
\phi\left(t\right)\simeq 
 {\phi_{i} \over a(t)^{3/2}}\cos(m_\phi t +\delta_\phi)~,\label{eq: phi solution gen no defined phi}
\end{equation}
where we have defined the initial field amplitude, $\phi_{i}$, and phase, $\delta_\phi$. The energy density of the oscillating field thus evolves, after a few oscillations, as:
\begin{eqnarray}\label{eq:phi density}
\rho_\phi \simeq  {1\over 2}{m^2\phi_{i}^2\over a^3}~,
\end{eqnarray}
which corresponds to the behavior of non-relativistic matter.

We may therefore consider an oscillating scalar field as a plausible dark matter candidate, provided that it is stable and yields the correct present abundance. In general, the field will begin to oscillate after inflation when $m_{\phi}\simeq H$. If we consider that the field only acquires mass through the Higgs mechanism, its mass vanishes before electroweak symmetry breaking and, consequently, $H>m_{\phi}$ and the field is overdamped, such that its amplitude remains approximately constant. After
the electroweak phase transition at temperatures around $100$ $\mathrm{GeV}$, the field acquires a mass that eventually becomes larger than the Hubble parameter. The field then becomes underdamped and begins to oscillate as obtained above. This will generically occur during the radiation-dominated epoch, where the Hubble expansion rate is given by:
\begin{equation}
H=\frac{\pi}{\sqrt{90}}\sqrt{g_{*}}\frac{T^{2}}{M_{Pl}}~,\label{eq:Hubble constant radiation}
\end{equation}
where $M_{Pl}$ is the reduced mass Planck, $M_{Pl}=1/\sqrt{8\pi G}$, $T$ is the cosmic temperature and $g_{*}=N_{B}+ (7/8)N_{F}$ is the total number of relativistic degrees of freedom, including $N_{B}$ and $N_{F}$ bosonic and fermionic degrees of freedom, respectively.

From Eq. (\ref{eq:phi density}), we may define an effective number density of $\phi$ particles in the oscillating scalar condensate:
\begin{equation}
n_{\phi}=\frac{\rho_{\phi}}{m_{\phi}}=\frac{m_{\phi}\phi_{i}^{2}}{2a^{3}}~.\label{eq:phi number density}
\end{equation}
The total entropy density of radiation in the early Universe is given by:
\begin{equation}
s=\frac{2\pi^{2}}{45}g_{*S}T^{3}~,\label{eq:entropy density}
\end{equation}
where $g_{*S}=N_{B}+ \frac{3}{4} N_{F}$ is the effective number of relativistic degrees of freedom contributing to the entropy. Using Eqs. (\ref{eq:phi number density}) and (\ref{eq:entropy density}), it is easy to see that the number of particles in a comoving volume is:
\begin{equation}
\frac{n_{\phi}}{s}=\frac{m_{\phi}\phi_{i}^{2}/(2a^3)}{\frac{2\pi^{2}}{45}\,g_{*S}\,T^{3}}=\mathrm{const.}\label{eq:n/s conserved quantity}
\end{equation}
This is a conserved quantity since, due to the entropy conservation, $S=sa^{3}$ remains constant throughout the history of the Universe. 

We consider now two separate cases, since the field only acquires its mass after the electroweak phase transition at $T_{EW}\sim 100$ GeV. If, on the one hand, the field mass is smaller than the Hubble rate $H_{EW}= \pi/\sqrt{90} g_* T_{EW}^2/M_P \sim 10^{-5} $ eV, with $g_*\sim 100$ \footnote{Note that the electroweak phase transition is not instantaneous, and both the temperature and the number of relativistic species vary during the phase transition. This simplified approach to consider a given temperature and $g_*$, it gives nevertheless a sufficiently good approximation for determining the main properties of the dark matter field.}, the field will only start to oscillate after the phase transition. If, on the other hand, $m_\phi\gtrsim H_{EW}$, oscillations start as soon as the Higgs field acquires its vacuum expectation value, which we take approximately to be at $T_{EW}$.

In the first case, for $m_\phi \lesssim H_{EW}$, the temperature at which $m_\phi= H$ is given by:
\begin{equation}
T=\left(\frac{90}{\pi^2}\right)^{1/4}g_*^{-1/4}\sqrt{M_{Pl}m_{\phi}}~,\label{eq:temperature oscillations begin}
\end{equation}
which is valid for temperatures below $T_{EW}$. Introducing this temperature into the relation Eq. (\ref{eq:n/s conserved quantity}), we get:
\begin{equation}
\frac{n_{\phi}}{s}={1\over8}\left({90\over \pi^2}\right)^{1/4} g_*^{-1/4} {\phi_{i}^2\over \sqrt{m_\phi M_{Pl}^3}}~,\label{eq:n/s conserved quantity step 2}
\end{equation}
where we have taken $g_*=g_{*S}$ when field oscillations begin. We may then use this to compute the present dark matter abundance, $\Omega_{\phi,0}$, defined as:
\begin{eqnarray}
\Omega_{\phi,0}&\equiv&\frac{\rho_{\phi,0}}{\rho_{c,0}}=\frac{m_{\phi}}{3H_{0}^{2}M_{Pl}^{2}}\left(\frac{n_{\phi}}{s}\right)s_0\nonumber\\
& \simeq & {1\over 6}\left({\pi^2\over 90}\right)^{3/4}{g_{*S0}\over g_*^{1/4}}{T_0^3 m_\phi^{1/2}\phi_{i}^2\over H_0^2 M_{Pl}^{7/2}}~,\qquad m_\phi < H_{EW}~,\label{eq:omega 0 final 1}
\end{eqnarray}
where $H_{0}\simeq1.45\times10^{-33}$ eV is the present Hubble parameter, $T_{0}\simeq2.58\times10^{-4}$ eV is the present CMB temperature and  $g_{*S0}\simeq 3.91$.

For the case where the field starts oscillating immediately after the electroweak phase transition, for $m_\phi\gtrsim H_{EW}$, we take the temperature at the beginning of field oscillations to be $T_{EW}$ and, following the same steps as for the previous case, we obtain:
\begin{eqnarray}
\Omega_{\phi,0}\simeq  {1\over6}{g_{*S0}\over g_*} \left({T_0\over T_{EW}}\right)^3 {m_\phi^2\phi_{i}^2\over H_0^2 M_P^2}~, \qquad m_\phi > H_{EW}~.
\label{eq:omega 0 final 2}
\end{eqnarray}
Then, assuming that the field accounts for all of the present dark matter abundance, $\Omega_{\phi,0}\simeq 0.26$ \cite{Ade:2015xua}, we obtain the following relations between the field mass and its initial amplitude:
\begin{eqnarray}
m_\phi\simeq \begin{cases}
3\times 10^{-5} \left({g_*\over 100}\right)^{1/2}\left({\phi_{i}\over 10^{13}\ \mathrm{GeV}}\right)^{-4}\ \mathrm{eV}~, & m_\phi < H_{EW}\\
2\times 10^{-5}\left({g_*\over 100}\right)^{1/2} \left({\phi_{i}\over 10^{13}\ \mathrm{GeV}}\right)^{-1}\ \mathrm{eV}~, & m_\phi > H_{EW}
\end{cases}~.
\label{eq:field mass}
\end{eqnarray}
%



\section{Inflation and initial conditions for the scalar field}  \label{sec: constraints Infl}

In the previous section we have determined the values of the field mass that may account for the present dark matter abundance as a function of its initial oscillation amplitude. As has been previously observed in the literature \cite{Enqvist:2014zqa}, the initial conditions for scalar field oscillations in the post-inflationary Universe are set by the inflationary dynamics itself, depending on whether the field mass is greater or smaller than the inflationary Hubble parameter, $H_{inf}$. 

We have assumed above that the dark matter field $\phi$ acquires mass through the Higgs mechanism, such that its mass during inflation would depend on the inflationary dynamics of the Higgs field itself. As we will see below, we will be interested in extremely small couplings between $\phi$ and the Higgs field of order $v/M_{Pl}\sim 10^{-16}$, where $v=246$ GeV is the Higgs vacuum expectation value (vev). This implies that the field mass during inflation will be at most of the order of the electroweak scale unless the Higgs field acquires super-planckian values. The dark matter field would thus be light during inflation, and consequently exhibit de-Sitter fluctuations of order $H_{inf}/2\pi$ on super-horizon scales. This would be phenomenologically unacceptable, since this would lead to large inhomogeneities in the dark matter density that would lead to sizeable cold dark matter isocurvature modes in the CMB spectrum as e.g.~for the case of axions (see e.g.~ Ref. \cite{Marsh:2014qoa}).

The Higgs field need not, however, be the unique source of mass for the dark matter field. In fact, in most extensions of the Standard Model with additional scalar fields, the latter typically acquire masses of the order of the Hubble parameter during inflation. This is, for example, the case of supergravity models, where the scalar potential involves terms of the form $V(\phi)\sim e^{K(\phi)/M_{Pl}^2} \mu^4+\dots\sim \mu^4 +\mu^4 |\phi|^2/M_{Pl}^2+\ldots$, where $\mu^4$ is the inflationary energy density, for canonical forms of the K\"ahler potential. This results in field masses of order $\mu^2 /M_P \sim H_{inf}$, which is the origin of the so-called ``eta-problem" found in supergravity/string inflationary scenarios (see e.g. Ref. \cite{Baumann:2014nda}). 

From a more general effective field theory point of view, we may argue that, even if there are no direct renormalizable interactions between the dark matter and the inflaton scalar fields, gravitational interactions may induce non-renormalizable terms of the form:
\begin{equation}
\mathcal{L}_{int}={c\over 2}\frac{\phi^{2}V(\chi)}{M_{pl}^{2}}~,\label{eq: Lint inflaton phi}
\end{equation}
where $\chi$ is the inflaton field and $c$ is a dimensionless parameter. This leads to a contribution to the field mass $m_\phi \sim c H_{inf}$ during inflation that vanishes in the post-inflationary era, assuming that $V(\chi)=0$ in the ground state. The magnitude (and sign) of this mass cannot be determined in the absence of a UV-complete description of the theory, but in the absence of fine-tuning we expect $|c|\sim \mathcal{O}(1)$, and we will focus on the case $c>0$ where the minimum of the $\phi$ potential lies at the origin. Note that the reheating period may have some effects on the dynamics of the fields, but it does not affect our scenario, as shown in \ref{app}.

A massive field with $m_\phi\sim H_{inf}$ will nevertheless exhibit quantum fluctuations that get stretched and amplified by expansion during the quasi-de Sitter inflationary phase.
For $m_{\phi}/H_{inf}<3/2$, the amplitude of each Fourier mode with comoving momentum $k$ is given by \cite{Riotto:2002yw}:
\begin{equation}
\left|\delta\phi_{k}\right|\simeq\frac{H_{inf}}{\sqrt{2k^{3}}}\left(\frac{k}{aH_{inf}}\right)^{\frac{3}{2}-\nu_{\phi}}~,\label{eq: fluctuations massive field}
\end{equation}
where $\nu_{\phi}=\left(\frac{9}{4}-\frac{m_{\phi}^{2}}{H_{inf}^{2}}\right)^{1/2}$. Notice that fluctuations do not ``freeze" on super-horizon scales, for $k<aH_{inf}$, unless $\nu_\phi\simeq 3/2$, i.e.~unless the field is very light. Instead, for a massive field, fluctuations are exponentially damped as inflation proceeds. The homogeneous field component can be obtained by integrating over all super-horizon modes, yielding for the variance of the scalar field:
\begin{equation}
\left\langle \phi^{2}\right\rangle =\left(\frac{H_{inf}}{2\pi}\right)^{2}\frac{\left(1-\left(e^{-N_{e}}\right)^{3-2\nu_{\phi}}\right)}{3-2\nu_{\phi}}\simeq\frac{1}{3-2\nu_{\phi}}\left({H_{inf}\over 2\pi}\right)^2~,
 \label{eq: <phi^2>}
\end{equation}
where, in the last step, we have taken $e^{-N_e}\ll 1$ for $N_e=50-60$ e-folds of inflation. Notice that the field variance becomes constant, even though each super-horizon Fourier mode is continuously damped. This is associated with the fact that there are always modes $k\sim aH_{inf}$ that give a significant contribution to the variance. 

As mentioned above, fluctuations in the dark matter scalar field will lead to isocurvature perturbations in the CMB spectrum, which are then given by \cite{Riotto:2002yw}:
\begin{eqnarray}
\mathcal{P}_{I}\left(k\right)\equiv\left\langle \left(2\frac{\delta\phi_{i}}{\phi_{i}}\right)^{2}\right\rangle \simeq\frac{2\pi^{2}}{k^{3}}\left(\frac{k}{aH_{inf}}\right)^{3-2\nu_{\phi}}\left(3-2\nu_{\phi}\right)~. \label{eq:power spectrum iso perturbations}
\end{eqnarray}
For $m_\phi/H_{inf}> 3/2$, the fluctuations are more suppressed, and given approximately by \cite{Riotto:2002yw}:
\begin{equation}
\left|\delta\phi_{k}\right|^{2}\simeq\left(\frac{H_{inf}}{2\pi}\right)^{2}\left(\frac{H_{inf}}{m_{\phi}}\right)\frac{2\pi^{2}}{\left(aH_{inf}\right)^{3}}~.\label{eq:flutuacoes para m/H maior 9/4}
\end{equation}
Integrating over k, at the end of inflation, we get for the field variance:
\begin{eqnarray}
\left\langle \phi^{2}\right\rangle \simeq\left(\frac{H_{inf}}{2\pi}\right)^{2}\left(\frac{H_{inf}}{m_{\phi}}\right)\left(1-e^{-3N_{e}}\right)\simeq\left(\frac{H_{inf}}{2\pi}\right)^{2}\left(\frac{H_{inf}}{m_{\phi}}\right)~,\label{eq: phi2 m/H menor 9/4}
\end{eqnarray}
and the isocurvature power spectrum is:
\begin{equation}
\mathcal{P}_{I}(k)\simeq\frac{2\pi^{2}}{\left(aH_{inf}\right)^{3}}.\label{eq: iso power maior 9/4}
\end{equation}
Let us focus on the case $m_\phi\sim H_{inf}$ (with real values of $\nu_\phi$), and determine the minimum field mass during inflation that leads to CDM isocurvature perturbations compatible with observations. For this we consider the dimensionless power spectrum:
\begin{equation}
\Delta_{I}^{2}\equiv\frac{k^{3}}{2\pi^{2}}\mathcal{P}_{I}(k)=\left(3-2\nu_{\phi}\right)\left(\frac{k}{aH_{inf}}\right)^{3-2\nu_{\phi}} \label{eq:delta^2phi}~.
\end{equation}
Notice that the comoving scales that are relevant for CMB perturbations have left the horizon 50-60 e-folds before the end of inflation, such that at the end of inflation $k/aH_{inf}\simeq e^{-N_e}\ll 1$. Isocurvature modes are then measured in terms of the ratio:
\begin{equation}
\beta_{iso}(k)=\frac{\Delta_{I}^{2}\left(k\right)}{\Delta_\mathcal{R}^{2}\left(k\right)+\Delta_{I}^{2}\left(k\right)},\label{eq:beta iso def}
\end{equation}
where $\Delta_{R}^{2}\simeq 2.2\times 10^{-9}$ is the amplitude of the adiabatic curvature perturbation spectrum generated by the inflaton field $\chi$. Note that the dark matter field is sub-dominant during inflation, so that its fluctuations do not induce perturbations in the space-time curvature. CDM isocurvature modes and adiabatic modes will be uncorrelated, since fluctuations in $\phi$ and $\chi$ are independent. The Planck collaboration places an upper bound on uncorrelated CDM isocurvature perturbations $\beta_{iso}\left(k_{mid}\right)<0.037$ for $k_{mid}=0.050$ $\mathrm{Mpc}^{-1}$ \cite{Ade:2015lrj}. Using the above results, this yields $\nu_\phi \lesssim 1.3$ for 55 e-folds of inflation, implying $m_\phi\gtrsim 0.75 H_{inf}$.

This lower bound on the dark matter field mass during inflation allows us to place an upper bound on the variance of the field at the end of inflation of $\langle \phi^2\rangle \lesssim 0.25^{2}\,H_{inf}^{2}$, and for masses of the order of the Hubble parameter during inflation we have $\langle \phi^2\rangle =\alpha^{2} H_{inf}^{2}$, with $\alpha \simeq 0.1-0.25$. This variance will set the average amplitude of the field at the end of inflation, and since the field remains overdamped until after the electroweak phase transition, we take the initial amplitude for field oscillations in the post-inflationary era to be in this range, i.e.:
\begin{equation}
\phi_{i} \simeq \alpha H_{inf}~, \qquad \alpha\simeq 0.1-0.25~.
\end{equation}
We may express the Hubble parameter during inflation in terms of the tensor-to-scalar ratio, $r= \Delta_t^2/\Delta_\mathcal{R}^2$, since the amplitude of the primordial gravitational wave spectrum is a direct probe of the inflationary energy scale, with:
\begin{equation}
H_{inf} = {\pi\over \sqrt{2}}\sqrt{\Delta_\mathcal{R}^2}M_P\sqrt{r} \simeq 2.5\times 10^{13}\left({r\over 0.01}\right)^{1/2}\ \mathrm{GeV}~.
\end{equation}
Replacing this into Eq.~(\ref{eq:field mass}), we can obtain a relation between the dark matter field mass and the tensor-to-scalar ratio:
\begin{eqnarray}
m_\phi\simeq 2\times 10^{-5}  \left({g_*\over 100}\right)^{1/2} \times \begin{cases}
(\alpha/0.25)^{-4}(r/0.03)^{-2}~, & m_\phi < H_{EW}\\
(\alpha/0.25)^{-1}(r/ 0.03)^{-1/2}~, & m_\phi > H_{EW}
\end{cases}\ \mathrm{eV}~.
\label{eq:field mass_r}
\end{eqnarray}

\begin{figure}[h]
\begin{centering}
\includegraphics[scale=1.2]{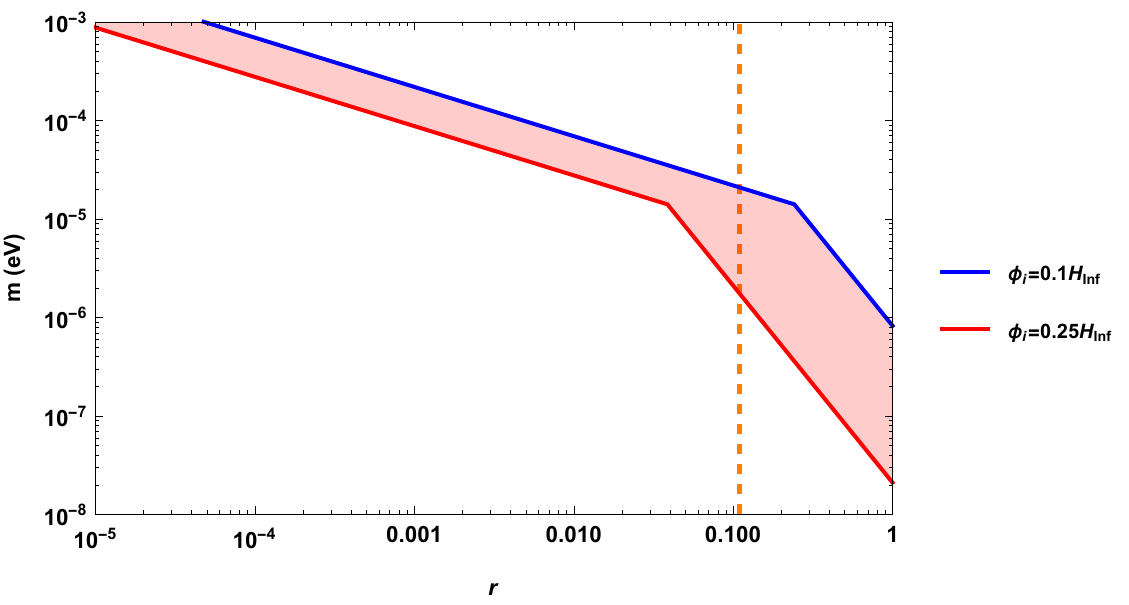}
\par\end{centering}
\caption{Relation between the dark matter field mass and the tensor-to-scalar ratio for an initial field amplitude $\phi_{i}=\alpha H_{inf}$ set by inflationary de Sitter fluctuations, with $\alpha=0.1-0.25$, corresponding to field masses during inflation $m_\phi\simeq (0.75-1.5)H_{inf}$. The dashed line gives the upper bound on the tensor-to-scalar ratio set by the Planck collaboration at $95\%$ C.L.~\cite{Ade:2015lrj}.}
 \label{fig:scalar-to-tensor-ratio}
\end{figure}

This relation is illustrated in Fig.~\ref{fig:scalar-to-tensor-ratio}, where one can see that the upper bound $r<0.11$ set by the Planck collaboration at $95\%$ C.L.~\cite{Ade:2015lrj} leads to a lower bound on the field mass $m_\phi\gtrsim 10^{-6}-10^{-5}$ eV for $\alpha\simeq 0.1-0.25$.

In the next sections we discuss possible scenarios that may lead to field masses of this order through the electroweak Higgs mechanism.



\section{ Non-renormalizable interactions between the dark matter and Higgs fields}  \label{sec:Dark matter coupled to Higgs via non-renormalizable operators}

The present dark matter abundance and the initial conditions set by the inflationary dynamics require very small scalar field masses, unless the tensor-to-scalar ratio is very suppressed. If the dark matter field were to couple directly to the Higgs field, via renormalizable operators, we would expect a mass not much below e.g.~the electron mass, unless the coupling is unnaturally small. However, we may envisage theories where such couplings are not present and, for instance, the dark matter field belongs to a {\it hidden} sector that is sequestered from the {\it visible} sector, which contains the Standard Model fields and in particular the Higgs boson. The absence of bare mass for the dark matter field could, for example, be motivated by a conformal symmetry or scale invariance in the hidden sector. The two sectors may, nevertheless, be indirectly coupled through heavy messenger fields that are e.g.~charged under the gauge symmetries of both sectors. Even in the absence of such messengers, the two sectors will be coupled through gravity. Taking a strictly effective field theory point of view, we may consider that the conformal symmetry/scale invariance in the hidden sector is broken only by non-renormalizable terms suppressed by a mass scale $M$ that corresponds to the messenger mass, or $M_P$ in the case of gravity-mediated interactions.

Imposing a  $\mathbb{Z}_{2}$  reflection symmetry on the dark matter field, i.e. the invariance of the Lagrangian under $\phi\rightarrow-\phi$, 
we can ensure the stability of the field, since linear terms that allow for its decay are thus forbidden. In this case, the lowest-order non-renormalizable operator involving the dark matter and the Higgs field is of dimension-6 and takes the form:
\begin{equation}
\mathcal{L}_{int}={a_6^2\over2}\mathrm{\left|h\right|}^{4}\frac{\phi^{2}}{M^{2}}~,\label{eq:L_int Higgs 1}
\end{equation}
where $\mathrm{a_6}$ is a dimensionless parameter, which we expect to be $\mathcal{O}(1)$. It is then easy to see that, after electroweak symmetry breaking when the Higgs field acquires its vev, the dark matter field acquires a mass:%

\begin{eqnarray}
m_{\phi}=a_6\frac{v^2}{M}\sim2.5\times10^{-5}a_6\left({M\over M_P}\right)^{-1}\quad\mathrm{eV}~.\label{eq: mass term vev}
\end{eqnarray}
This implies that one obtains a dark matter field mass in the range required by the observed abundance for Planck-suppressed interactions, assuming inflation occurs close to the GUT scale and the tensor-to-scalar ratio is not too suppressed ($r\gtrsim 10^{-3}$). If inflation occurs at an energy scale somewhat below the GUT scale, leading to lower values of the tensor-to-scalar ratio, the observed dark matter abundance would require larger values of $m_\phi$, and these could equally be motivated e.g.~by messenger masses at or below the GUT scale.

The case of Planck-suppressed interactions seems, however, rather special, since the hierarchy between the electroweak and Planck scales leads to a similar hierarchy between the dark matter and Higgs masses. This motivates going beyond the effective field theory perspective and finding a concrete and well-motivated scenario where this hierarchy is naturally obtained, as we describe in the next section.



\section{Scalar field dark matter in warped extra-dimensions}  \label{sec:Scalar field dark matter in a higher dimensional theory}

A concrete realization of the hierarchy mentioned in the previous section in the context of an effective field theory can be found in the Randall-Sundrum (RS) scenario for warped extra-dimensions \cite{Randall:1999ee}. The RS construction considers a model with one additional spatial dimension with a warped geometry, and which is compactified in an orbifold $S^{1}/\mathbb{Z}_{2}$. This geometry can be viewed as an effective dimensional reduction of brane-models in 10/11-dimensional string/M-theory, where brane tensions are responsible for warping the geometry in the directions transverse to their world-volume.

 The bulk geometry is a slice of anti-de Sitter space ($AdS_{5}$), where the metric corresponds to the warped product:
\begin{equation}
ds^{2}=e^{-2\sigma\left(y\right)}g_{\mu\nu}dx^{\mu}dx^{\nu}+dy^2,\label{eq:metrica ads}
\end{equation}
where $y\in [-L, L]$ is the radial coordinate along the extra-dimension, for $L=\pi r_c$ with $r_c$ denoting the compactification radius, and  $\mu,\nu=0,1,2,3$. The warp factor is given in terms of the linear function $\sigma(y)=k|y|$, where $k$ is the bulk AdS curvature. The orbifold symmetry has two fixed points at $y=0$ and $y=L$, where branes of opposite tension reside. Einstein's equations then require a particular relation between the brane tensions and the (negative) bulk cosmological constant.

In their original proposal, Randall and Sundrum observed that if the Standard Model fields, in particular the Higgs boson, were confined to the brane at $y=L$, the Higgs mass and its expectation value would be exponentially suppressed with respect to the fundamental mass scale in the construction, which is taken to be the Planck scale. One then naturally obtains a large hierarchy $e^{-kL}\simeq v/M_P\simeq 10^{-16}$ between the electroweak and gravitational scales for a relatively small extra-dimension, providing a simple solution to the well-known gauge hierarchy problem. The Randall-Sundrum model has, since its original proposal, been the object of several analyses and extensions, including e.g.~scenarios where the Standard Model gauge and fermion fields reside in the bulk \cite{Gherghetta:2000qt,Bertolami:2007dt}, which could explain e.g.~the measured fermion mass hierarchy.

Here we will show that the required hierarchy between the dark matter and Higgs field masses can also be naturally obtained within the RS construction, if the dark matter field corresponds to the zero-mode of a bulk scalar field that is coupled to the Higgs field on the ``visible" or ``infrared" brane at $y=L$. We start with the following five-dimensional action for the Higgs and dark matter sectors:
\begin{eqnarray} \label{eq:S Higgs scalar DM}
S & = & \int d^{4}x\int dy\sqrt{-G}\,\left[\frac{1}{2}G^{MN}\partial_{M}\Phi\partial_{N}\Phi-\frac{1}{2}M_{\Phi}^{2}\Phi^{2}\right.\nonumber \\
 & + & \left.\delta\left(y-L\right)\left(G^{MN}\partial_{M}h^{\dagger}\partial_{N}h-V\left(h\right)+\frac{1}{2}g_{5}^{2}\Phi^{2}h^{2}\right)\right]~, 
\end{eqnarray}
where $G_{MN}$ is the five dimensional metric, $\Phi$ is the 5-dimensional scalar field that includes the dark matter field as its zero-mode and $g_{5}$ is the five-dimensional coupling between the Higgs and bulk scalars on the visible brane. In general, we may include a bare mass term for the scalar field, that is even under the $\mathbb{Z}_{2}$ orbifold symmetry and can be parametrized as \cite{Gherghetta:2000qt}:
\begin{equation}
M_{\Phi}^{2}=ak^{2}+b\sigma''~,\label{eq:phi mass param}
\end{equation}
where $a$ and $b$ are dimensionless parameters yielding the bulk and boundary contributions to the field mass, such that:
\begin{equation}
\sigma'=\frac{d\sigma}{dy}=k\,\mathrm{sgn\left(y\right),}\label{eq:sigma y '}\qquad 
\sigma''=\frac{d^{2}\sigma}{dy^{2}}=2k\,\left[\delta\left(y\right)-\delta\left(y-L\right)\right].\label{eq: sigma y ''}
\end{equation}
From the five-dimensional action we may derive the equation of motion for the bulk scalar. Assuming that a perturbative approach is valid, we may first neglect the brane-localized interactions with the Higgs field, to obtain:
\begin{equation}
\frac{1}{\sqrt{-G}}\partial_{M}\left(\sqrt{-G}\,G^{MN}\partial_{N}\Phi\right)-M_{\Phi}^{2}\Phi=0~.\label{eq:scalar field eom}
\end{equation}
Using the metric Eq. (\ref{eq:metrica ads}), this can be written in the form:
\begin{equation}
\left[e^{2\sigma}g^{\mu\nu}\partial_{\mu}\partial_{\nu}+e^{4\sigma}\partial_{y}\left(e^{-4\sigma}\partial_{y}\right)-M_{\Phi}^{2}\right]\Phi\left(x^{\mu},y\right)=0~.\label{eq:2nd order diff eq for phi}
\end{equation}
We can then decompose the bulk scalar into a tower of Kaluza-Klein modes:
\begin{equation}
\Phi\left(x^{\mu},y\right)=\frac{1}{\sqrt{2L}}\sum_{n=0}^{\infty}\phi_n\left(x^{\mu}\right)f_{n}\left(y\right)~,\label{eq:KK modes}
\end{equation}
where the mode functions $f_{n}\left(y\right)$ satisfy the orthonormality condition:
\begin{equation}
\frac{1}{2L}\int_{-L}^{L}dy\,e^{-2\sigma}\,f_{n}\left(y\right)f_{m}\left(y\right)=\delta_{nm}~.\label{eq:orthonormal cond}
\end{equation}
Substituting into Eq. (\ref{eq:2nd order diff eq for phi}), we get:
\begin{equation}
\left[-e^{4\sigma}\partial_{y}\left(e^{-4\sigma}\partial_{y}\right)+M_{\Phi}^{2}\right]f_{n}=e^{2\sigma}\,m_{n}^{2}\,f_{n}.\label{eq: KK eigenmodes}
\end{equation}
This equation admits, in particular, a massless solution $f_0(y)$ with $m_n=0$ of the form:
\begin{equation}
f_{0}\left(y\right)=c_{1}^{0}\,e^{\left(2-\sqrt{4+a}\right)ky}+c_{2}^{0}\,e^{\left(2+\sqrt{4+a}\right)ky}~,\label{eq: massless mode wiht c1, c2}
\end{equation}
where $c_{1}^{0}$ and $c_{2}^{0}$ are constants, and the orbifold symmetry allows us to focus on the interval $y\in [0,L]$. This solution only exists for fields that are even under the orbifold $\mathbb{Z}_{2}$ symmetry \cite{Gherghetta:2000qt}, and which must satisfy the boundary conditions:
\begin{equation}
\left.\left(\frac{df_{n}}{dy}-b\sigma'f_{n}\right)\right|_{0,L}=0~.\label{eq:boundary condition}
\end{equation}
For the zero-mode, this implies $c_{1}^{0}=0$ and $b=2\pm\sqrt{4+a}$. Imposing the normalization condition (\ref{eq:orthonormal cond}), the zero-mode profile is then given by:
\begin{equation}
f_{0}\left(y\right)=\sqrt{\frac{2Lk\left(b-1\right)}{e^{2kL\left(b-1\right)}-1}}e^{bky}~.\label{eq:zero mode complete}
\end{equation}
We will now focus on the particular case $a=b=0$, for which the bulk scalar is scale invariant and the associated zero-mode function is flat:
\begin{equation}
f_{0}\left(y\right)\simeq\sqrt{2kL}~,\label{eq: f0}
\end{equation}
where we used that $e^{-kL}\sim v/M_P\ll 1$. If we now replace this mode-function into the brane-localized terms in the action, we find the following effective interaction between the zero-mode and the Higgs fields:
\begin{eqnarray} \label{eq:4D interaction}
\mathcal{L}_{\phi h}  ={1\over 2}g_5^2k e^{-2kL}\phi^2h^2~,
\end{eqnarray}
where we have rescaled the Higgs field $h\rightarrow e^{kL}h$ in order for it to have a canonically normalized kinetic term in four dimensions, and denoted the zero-mode field $\phi_0(x)\equiv \phi(x)$. Thus, noting that $g_5^2 k$ is a dimensionless quantity and that the AdS curvature $k\simeq M_P$ is the fundamental scale in the RS scenario, we expect the effective four-dimensional coupling between $\phi$ and the Higgs field to be:
\begin{eqnarray} \label{eq:4D interaction}
g =\sqrt{g_5^2k} e^{-2kL} \simeq \mathcal{O}(1)\times  {v\over M_P}\sim 10^{-16}~. 
\end{eqnarray}
Thus, assuming that the five-dimensional coupling has a natural value $g_5\sim k^{-1/2}$, we conclude that, after the electroweak symmetry is broken, the zero-mode of our bulk scalar acquires a mass $m_\phi \sim v^2/M_P\sim 10^{-5}$ eV, just like for the non-renormalizable interactions considered in the previous section. Of course the five-dimensional coupling may somewhat differ from the natural scale without much fine-tuning of the extra-dimensional model, but this shows that the zero-mode of a scale invariant bulk scalar in the RS model acquires a mass in the correct range to account for the dark matter in the Universe as an oscillating scalar field. The effect of a Planck-suppressed non-renormalizable operator is thus analogous to a renormalizable interaction in a higher-dimensional warped geometry.

One of the assumptions of the generic analysis we performed in the previous sections is that the dark matter field acquires a Hubble-scale mass during inflation from a non-renormalizable coupling to the inflaton field. This can be implemented within the RS construction if, for example, the inflaton field $\chi$ also lives on the visible brane and we consider brane-localized interactions of the form:
\begin{eqnarray}
S_{\Phi,\chi} = h_5 \int d^{4}x\int dy\sqrt{-G}\, \delta\left(y-L\right)\,\Phi^{2}\,V\left(\chi\right)~.\label{eq:S inf total}
\end{eqnarray}
Since, by dimensional analysis, $h_5\sim k^{-3}\sim M_P^{-3}$, it is easy to conclude that the effective four-dimensional coupling between the inflaton and the dark matter zero-mode field is of the form $\phi^2 V(\chi)/ k^2\sim \phi^2 V(\chi)/M_P^2$, thus naturally yielding a Hubble-scale mass for the dark matter field during inflation which vanishes in the post-inflationary era.

Within the RS construction, the remaining KK modes of the bulk scalar field could, in principle, also contribute to the dark matter density if they oscillate with a sufficiently large amplitude after inflation. Although we will not analyze the properties of these modes in detail, referring the reader to existing discussions in the literature as e.g. Ref. \cite{Gherghetta:2000qt}, we must ensure that they will not {\it overcontribute} to the dark matter density. In particular, for $n>1$, the mass spectrum for an even field with $a=b=0$ is given by:
\begin{eqnarray}
m_n\simeq \left(n+{1\over 4}\right)\pi k e^{-kL}~,
\end{eqnarray}
such that the lowest KK masses lie at the TeV scale and are, hence, much heavier than the zero-mode. This could lead to an overabundance of dark matter, but we note that the KK mode functions are given approximately by:
\begin{eqnarray}
f_n(y)\simeq\sqrt{2kL}e^{k(2y-L)}{J_2\left(m_ne^{ky}/k\right)\over J_2\left(m_ne^{kL}/k\right)}~.
\end{eqnarray}
This implies that their coupling to fields on the visible brane is exponentially larger than the coupling of the zero-mode, with $f_n(L)=e^{kL}f_0(L)$, and consequently that their mass during inflation is necessarily much larger than the inflationary Hubble scale. Since these may actually be super-planckian, it is not possible to study their dynamics during inflation as for the case of the zero-mode, but this analysis nevertheless shows that we do not expect the KK modes of the bulk scalar to develop large expectation values during inflation and hence oscillate with a large amplitude in the post-inflationary eras. However, those heavy modes might decay rather quickly through gravitational coupling and then we expect that indeed only the zero-mode contributes significantly to the present abundance of non-relativistic matter.

\begin{figure}[htbp]
\begin{centering}
\includegraphics[scale=0.9]{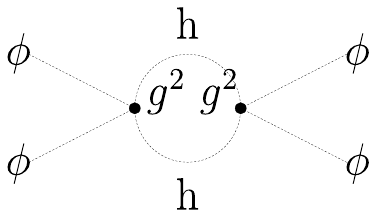}
\par\end{centering}

\caption{Feynman diagram inducing a dark matter self-coupling at 1-loop order.}
\label{fig:feyn diagram}

\end{figure}

Finally, in our previous dynamical analysis of the oscillating dark matter field, we have neglected the effects of any field self-interactions. Although the assumption of a bulk conformal symmetry for the five-dimensional scalar field implies that no bare self-interaction terms exist, we must take into account that this symmetry is broken on the visible brane by the interactions with the Higgs field (as well as the inflaton but this does not affect the post-inflationary dynamics). This generates, in particular, a quartic coupling for the dark matter zero-mode field through radiative corrections. At 1-loop, these corrections correspond to the diagram in Fig.~\ref{fig:feyn diagram}, which up to numerical factors and the usual logarithms generates a quartic-self coupling $\lambda \sim g^4$. 

Since after inflation the field has a large expectation value $\phi_{i} \sim \alpha H_{inf}$, these self-interactions yield a contribution to the dark matter field mass $\Delta m_\phi^2 \sim \lambda \phi_{i}^2 \sim g^4 H_{inf}^2$. It is easy to check that $\Delta m_\phi^2 / m_\phi^2 \sim H_{inf}^2/ M_P^2\ll 1$, so that we may safely neglect the effect of these self-interactions on the dynamics of the dark matter field.



\section{Conclusions}  \label{concl}

In this work, we have analyzed the possibility of an oscillating scalar field, $\phi$, which acquires mass through the Higgs mechanism, accounting for the observed dark matter abundance in the Universe. 

We have argued that the field acquires a large expectation value during inflation, just below the inflationary Hubble parameter, due to quasi-de Sitter quantum fluctuations, which sets the average initial amplitude of the field oscillations. Despite the large overall variance of the field, its fluctuations on the presently observable CMB scales are exponentially damped, since scalar fields generically acquire Hubble-scale masses through gravitationally-induced couplings to the inflaton. Thus, in contrast with e.g. axion fields, which remain massless during inflation and acquire large fluctuations on all super-horizon scales, inflationary fluctuations would not generate significant matter isocurvature perturbations in the present scenario.

After inflation the field becomes massless and its amplitude remains approximately frozen until the electroweak phase transition, after which the field acquires a small mass through its coupling to the Higgs field. Eventually the field becomes underdamped and begins to oscillate about the origin, behaving as cold dark matter. We have also imposed a $\mathbb{Z}_2$ symmetry on the Lagrangian that ensures the stability of the dark matter field. The observed dark matter abundance then allows us to determine the field mass as a function of the inflationary Hubble scale, and the current Planck bound on the tensor-to-scalar ratio \cite{Ade:2015lrj} sets a lower bound on the field mass $m_\phi\gtrsim 10^{-6}-10^{-5}$ eV. 

If inflation occurs close to the GUT scale and the tensor-to-scalar ratio is within the range of planned CMB experiments, $r\gtrsim 10^{-3}$, we concluded that the mass of the scalar field must saturate the above bound, and we observed that this implies the approximate hierarchy between the dark matter mass and the Higgs expectation value:
\begin{eqnarray}
{m_\phi\over v}\sim {v\over M_P}~. 
\end{eqnarray}
We have then explored different scenarios where this hierarchy may be attained. A first, more generic, possibility is that the dark matter and Higgs fields are only coupled through Planck-suppressed gravitational interactions. We have, in addition, presented a more concrete possibility that the dark matter field is the zero-mode of a five-dimensional scalar field living in a warped extra-dimension. In the context of the Randall-Sundrum model, this mode acquires a mass through its coupling to the Higgs field, which is confined to the visible brane at the bottom of the warped throat. We have also shown that, within this scenario, this field acquires a Hubble-scale mass during inflation, as initially assumed, when the the inflaton is also localized on the visible brane.

A generic feature of these scenarios is that the dark matter field has no bare mass or self-interactions, which may be motivated by imposing a conformal symmetry that is broken only by the non-renormalizable or brane-localized interactions with the Higgs field after electroweak symmetry breaking. The resulting interactions between the dark matter and Higgs bosons are thus extremely suppressed, with an effective coupling $g\sim 10^{-16}$. On the one hand, this justifies neglecting any dissipative effects in the dynamics of the oscillating scalar field, which could e.g.~lead to its evaporation and subsequent thermalization as considered in \cite{Tenkanen:2015nxa}. On the other hand, this will make its detection extremely difficult. Although we will leave a detailed analysis of possible experimental signatures for a future publication, we expect that such a dark matter candidate will evade detection in ongoing or even planned experiments involving nuclear recoil. 

The dark matter field exhibits, however, some similarities with axions and other axion-like particles, namely in terms of its small mass and couplings to known particles. This suggests that it may be possible to probe the existence and properties of the proposed dark matter scalar field with experiments analogous to those employed in the search of axion-like particles, or even using similar indirect astrophysical signatures. The nature and interaction structure of these fields are, nevertheless, sufficiently different that one may hope to distinguish them experimentally. 

This work shows that dark matter, as the known particles in the Standard Model, may acquire mass through the Higgs mechanism, despite its hidden/sequestered nature. It shows, in addition, that the Higgs portal can offer alternative dark matter candidates to the thermal WIMPs typically considered in the literature, and that the properties of dark matter can also be used to probe the mechanism behind inflation in the early Universe. We thus hope that our work motivates further exploration of the different possibilities presented and other potentially related scenarios.

\vspace{0.5cm} 

\noindent
{ \bf Acknowledgements}

\noindent
The work of C.C.~is supported by the Funda\c{c}\~{a}o para a Ci\^{e}ncia e a Tecnologia (FCT) under the grant
PD/BI/106012/2014. J.G.R.~is supported by the FCT grant SFRH/BPD/85969/2012 and partially by the H2020-MSCA-RISE-2015 Grant No. StronGrHEP-690904, and by the CIDMA project UID/MAT/04106/2013.

\appendix
\section{Effects of the reheating period} \label{app}

Despite the mass of the dark matter field vanishing in the radiation-era before electroweak symmetry breaking, one must check whether the reheating period can affect the results presented in Section 3. During this period,
\begin{equation}
{m_\phi^2\over H^2} = 3c {V(\chi)\over V(\chi)+{1\over2}\dot\chi^2+\rho_R}~. 
\end{equation}
Since during inflation we have $m_\phi^2/H^2 \simeq 3c \sim 1$, the dark matter field will be overdamped (or at most critically damped), $m_\phi \lesssim H$, in the post-inflationary eras, as we mentioned above, and we do not expect the dark matter field to oscillate during reheating. Nevertheless, before the radiation era, where $m_\phi^2/H^2\sim V(\chi)/\rho_R \ll 1$, there may be a period of inflaton matter-domination, where the latter oscillates about the minimum of its potential but has not yet decayed significantly. Neglecting $\rho_R$ in the equation above, we have $m_\phi^2/H^2 = 3c/2$, and it is easy to check that during this period the field amplitude decays as $t^\alpha$, where 
\begin{equation}
\alpha = {1\over2}\left(-1+\sqrt{1-{8\over3}c}\right)~.
\end{equation}
Thus, the field does not oscillate in this period for $c<3/8$, which corresponds to $m_\phi \lesssim 1.1 H_{inf}$ during inflation, which is compatible with the Planck bounds on CMB isocurvature modes. For example, for $c=1/3$, we have $\alpha=-1/3$, such that the field amplitude will decay as $a^{-1/2}$, implying that the inflaton-dominated matter era may last for a few e-folds without significantly decreasing the dark matter field amplitude. The duration of this era is, of course, model-dependent, and we note that there are scenarios where this era is, in fact, absent and the slow-roll regime is immediately followed by radiation-domination, such as when the inflaton has a quartic potential or in warm inflation scenarios (see e.g.~\cite{Bartrum:2013fia}).

The interactions between the inflaton and the $\phi$ field may also lead to the production of $\phi$-particles during reheating. We expect this to be a negligible process in general due to the non-renormalizable nature of the interactions, since close to the minimum of the inflaton potential at $\chi_0$, we have:
\begin{equation}
\mathcal{L}= c{V(\chi)\phi^2\over M_P^2} = {c\over 2} {m_\chi^2\over M_P^2} \chi^2\phi^2+\ldots\equiv g^2  \chi^2\phi^2+\ldots
\end{equation}
where $m_\chi^2= V''(\chi_0)$ is the inflaton mass at the minimum and we assumed a vanishing vacuum energy. If this mass coincides with the inflaton mass during inflation, $m_\chi^2= 3\eta H_{inf}^2$, taking into account the amplitude and tilt of the primordial curvature spectrum, this yields an effective coupling:
\begin{equation}
g^2 \sim 10^{-12} \left({\eta\over 0.01}\right)\left({r\over 0.01}\right)~.
\end{equation}
This coupling may be even more suppressed if the inflaton mass at the minimum is considerably smaller than its value during inflation. It is thus not hard to envisage scenarios where the inflaton couples more strongly to Standard Model particles such that only a negligible fraction of the inflaton's energy is converted into $\phi$-particles during reheating, ensuring a sufficiently long radiation-dominated era. 

One may nevertheless ask whether such particles could contribute to the present dark matter abundance. Due to their extremely small coupling to Standard Model particles (as we describe in the previous sections) and to the inflaton, $\phi$-particles never thermalize in the cosmic history, so that their final abundance is set entirely by their initial abundance, i.e.~by the inflaton decay $\chi\rightarrow \phi\phi$. Their initial number density is thus (following Ref. \cite{Dev:2013yza}):
\begin{equation}
n_{\phi i}=B_\phi n_\chi =2 B_\phi {\pi^2\,g_*\over 30}{T_R^4\over m_\chi}~,
\end{equation}
where $T_R$ is the reheating temperature, assuming instantaneous reheating in the worst-case scenario, and $B_\phi$ is the branching ratio of inflaton decays into dark matter particles. The latter may be relativistic when produced if the $\phi$ mass is already considerably smaller than the inflaton mass, such that they have initial momentum $p_{\phi i}\sim m_\chi/2$. As they are always decoupled from the cosmic plasma, their momentum simply redshifts with expansion by a factor $e^{-N_e}$, where $N_e$ denotes the number of e-folds of expansion after inflation. Taking $m_\phi \sim \sqrt{\eta} H_{inf}\sim 10^{12}$ GeV and $N_e\simeq 60$, we obtain a present momentum  $p_{\phi0}\sim 10^{-5}$ eV, which is comparable to the present mass of the dark matter particles in the range of interest to our scenario. This means that the dark matter particles should only be mildly relativistic today, $E_{\phi 0}\sim m_\phi$. Then:
\begin{equation}
\Omega_{\phi0} \simeq {m_\phi n_{\phi0}\over 3H_0^2 M_P^2} = {m_\phi s_0\over 3H_0M_P^2} \left({n_{\phi i}\over s_i}\right)\simeq  0.01 B_\phi \left({m_\phi\over 10^{-5} \ \mathrm{eV}}\right)\left({T_R\over 10^{15}\ \mathrm{GeV}}\right)\left({m_\chi\over 10^{12}\ \mathrm{GeV}}\right)^{-1}~,
\end{equation}
where we have used that $n_\phi/s$ remains constant for a decoupled species. We thus see that, due to the smallness of their mass, $\phi$-particles from the inflaton decay generically give a negligible contribution to the present dark matter abundance, even if the branching ratio is not too suppressed.

In summary, we do not expect the reheating period to considerably modify our analysis in general, so that we may neglect its effects in computing the present dark matter abundance.


\end{document}